# Securing Untrusted Full-Duplex Relay Channels in the Presence of Multiple External Cluster-Based Eavesdroppers

Ahmed El Shafie[†], Asma Mabrouk[⋆], Kamel Tourki[∗], Naofal Al-Dhahir[†], Mazen O. Hasna[††]

[†]Electrical Engineering Dept., University of Texas at Dallas, (e-mail: {ahmed.elshafie, aldhahir}@utdallas.edu).
[⋆]HANA Research Lab, ENSI, Manouba University, (e-mail: mabroukasma89@gmail.com).
[∗]Mathematical and Algorithmic Sciences Lab, France Research Center, Huawei Technologies Co. Ltd, (e-mail: kamel.tourki@gmail.com).
[††]Electrical Engineering Dept., Qatar University (e-mail: hasna@qu.edu.qa).

*Abstract*—This letter investigates the physical layer security in a wireless cooperative network where communication is assisted by a full-duplex (FD) untrusted relay in the presence of multiple external eavesdroppers. A cluster-based colluding eavesdropping setting is considered, where illegitimate nodes with common interests are grouped in a cluster. In order to confuse the different eavesdropping clusters, we consider artificial-noise-aided beamforming at the source node. Moreover, FD relay jamming is adopted to improve the system's security. To maintain secure communications against the untrusted relay node, a FD destination jamming scheme is adopted. Our proposed scheme is designed based on the channel state information of the legitimate nodes only. Numerical results show that the optimal power allocation factor between data and artificial noise depends on the total number of antennas of the different colluding eavesdropping clusters.

## I. INTRODUCTION

Eavesdropping on network traffic is one of the most critical security attacks due to the broadcast nature of the wireless medium. Security techniques have traditionally been considered at the higher layers of communication networks based on cryptography. Although cryptography-based techniques improve the computational security of the system and are widely deployed, they do not provide information-theoretic security guarantees and they only rely on the fact that some mathematical problems are hard to solve at the eavesdroppers. To meet confidentiality requirements in complex and diverse scenarios, there has been a renewed interest in a multi-layer wireless security solution that exploits the time-varying nature of wireless channels at the physical layer, known as physical layer security (PLS), with cryptography protocols implemented at the application layer.

As proven in [1], based on PLS, a positive secrecy capacity is achieved only when the rate of the source-eavesdropper channel is worse than the rate of the source-destination channel. When the transmitter has multiple antennas, a widely used secrecy approach aims at reducing the eavesdropper's signal-to-interference-plus-noise ratio (SINR) by sending an artificial noise (AN), superimpose on the data signal, into the null space of the intended receiver's channel matrix [2] to ensure that only the eavesdropper's received signal will be degraded.

For the single-antenna source scenario, by using cooperative jamming (CJ) techniques [3], the helper nodes cooperatively inject AN signals independent of the confidential information signals to jam the eavesdropper nodes. Referred to as the destination-based jamming (DBJ) technique, the destination can perform CJ to degrade the channel quality of an internal eavesdropper, (such as an untrusted relay) [4], [5]. In fact, the relay nodes can be honest but curious nodes, hence, they are trusted at the service level but are untrusted at the information level. The authors of [6] proposed a joint DBJ and CJ scheme to reduce the secrecy outage probability of wireless cooperative systems under untrusted relaying.

The common theme of the aforementioned works is based on the fact that they mainly employ the half-duplex (HD) mode at the relay due to its low complexity. Recent progress in self-interference cancellation techniques paved the way to the use of full-duplex (FD) relaying scheme as a promising spectral efficient solution [7]. To improve the PLS, a hybrid HD/FD relay selection technique was proposed in [8]. Recently, FD-relaying with jamming was proposed to assist a single-antenna source in improving the system's secrecy capacity [9]. In fact, in this approach, the FD relay is designed to transmit a jamming signal to the eavesdropper while receiving the data from the source. To achieve a positive secrecy capacity, the authors of [10] considered an FD legitimate receiver that simultaneously receives its intended data and sends jamming signals to an FD untrusted relay. Unlike [10], we assume that the transmitting nodes are equipped with multiple antennas in the presence of external eavesdropping.

Generally, most PLS schemes are proposed against attacks by either an individual eavesdropper or independently-operating multiple eavesdroppers (i.e., non-colluding eavesdroppers). However, the subscribed users in a network (each with an arbitrary number of antennas) can construct collaborative eavesdropping clusters based on common interests (benefits) and operate in parallel as illegitimate nodes. In other words, there are many attacking systems and each one has its own nodes that work for its benefits. In this case, the non-colluding eavesdropping model may underestimate the adversary's power and fail to capture it. For an ideal attack strategy, when multiple eavesdroppers coexist, they may share their signal observations to make the eavesdropping attack more efficient (i.e., colluding eavesdroppers) [11]. Furthermore, for collaboration purposes, eavesdroppers can be classified by the type of their attacks, their geographic locations, or any beneficial incentives.

To the best of our knowledge, the cluster-based colluding eavesdropping model has not been studied in the literature before. In addition, we investigate the secrecy performance of the untrusted FD relay network in the presence of multiple external clusters of eavesdroppers. Our contributions are summarized as follows

- We consider the most general case of eavesdropping

This paper was made possible by NPRP grant number 8-627-2-260 from the Qatar National Research Fund (a member of Qatar Foundation). The statements made herein are solely the responsibility of the authors.

attacks and assume that multiple external eavesdropping nodes coexist with the FD untrusted relay. Furthermore, we assume that these eavesdropping nodes are organized into clusters and that attackers within the same cluster collude with each other to design joint attack.
- To achieve positive secrecy, we propose a new scheme to secure the source transmissions and derive the achievable secrecy rate. In our proposed scheme, the destination operates in the FD mode to be able to receive its intended data signal while jamming the untrusted relay. On the other hand, the multi-antenna nodes, the source and the relay, send jamming signals along the null space of the relay and the destination channel, respectively, to confound all eavesdroppers in the different clusters.

*Notation:* Unless otherwise stated, lower- and upper-case bold letters denote vectors and matrices, respectively. $\mathbf{I}_N$ denotes the identity matrix whose size is $N \times N$. $\mathbf{0}_{M \times N}$ denotes the zero matrix whose size is $M \times N$. $\mathbb{C}^{M \times N}$ denotes the set of all complex matrices of size $M \times N$. $(\cdot)^*$ denotes the Hermitian conjugate of a matrix. $\mathbf{0}_{M \times N}$ denotes the all-zero matrix with size $M \times N$. $\bar{\theta} = 1 - \theta$. blkd$=\{\cdot\}$ denotes a diagonal matrix with the enclosed elements as its diagonal elements. $[\cdot]^+ = \max\{\cdot, 0\}$ denotes the maximum between the value in the argument and zero.

## II. SYSTEM MODEL AND ASSUMPTIONS

We consider secure communication between a transmitter (Alice) and an FD receiver (Bob) through an FD amplify-and-forward (AF) untrusted relay node (Ray), in the presence of multiple eavesdroppers. Alice (A), Bob (B) and Ray (R) are assumed to be equipped with one receive and multiple transmit antennas. We denote the number of transmit antennas at node $j$ as $N_j$ where $j \in \{A, R, B\}$. We assume that the multiple alien/external eavesdroppers are located in the radio-frequency (RF) range of Alice, Bob and Ray. As shown in Fig. 1, we further assume that, according to their interdependent behaviors, the eavesdroppers are divided into $M$ different clusters/groups, $(\mathcal{G}_i, 1 \leq i \leq M)$. Eavesdroppers within the same cluster can jointly process their received information signals, which makes the security issue more challenging. We denote the total number of antennas of eavesdroppers belonging to the cluster $\mathcal{G}_i$ as $N_i^e$. We assume that the legitimate nodes **do not** have knowledge of the eavesdroppers' channel state information (CSI), This is reasonable since the eavesdroppers are passive nodes that try to access information of the legitimate nodes. The legitimate transmitters know the CSI of their own links to their respective receivers.

Alice transmits her information signal, $x$, to Bob through the FD relay. At the same time, Bob sends an AN signal, denoted by $\mathbf{z}_B$, to confuse Ray so that he is not able to decode Alice's data. The jamming signal from Bob, which is also embedded in Ray's forwarded signal, will only hurt the eavesdroppers. When Ray forwards the signal to Bob, the AN can be canceled out. This will secure the transmissions from both alien eavesdropping and from the untrusted relay. All the eavesdroppers are able to hear all transmissions. To secure her transmission, Alice will assign a portion of her power, $\theta_A$, for data and the rest, $\bar{\theta}_A$, to send an AN signal denoted by $\mathbf{z}_A$.

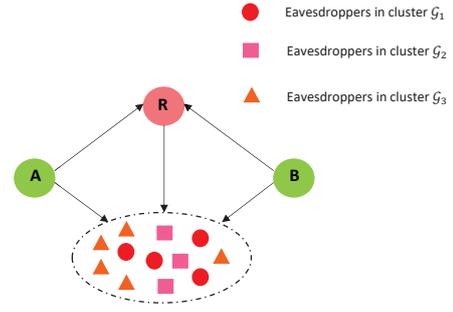

Fig. 1. A two-hop cooperative communication network with a wireless untrusted relay node. In the figure, we show an example of three eavesdropping clusters.

The AN signal from Alice is designed to be canceled at Ray to avoid forwarding the interference to Bob. Furthermore, Ray uses a part of his own power to send an AN signal, denoted by $\mathbf{z}_R$, orthogonal to the Ray-Bob channel vector. This AN signal will only jam the alien eavesdroppers. The data power ratio at R is denoted by $\theta_R$.

Time is partitioned into slots each with a duration of $T$. The channel has a bandwidth of $\mathcal{W}$. We assume a block-fading channel model where the channel coefficients remain constant during a time slot (coherence time) duration, but change independently from one time slot duration to another. We assume that $\mathbf{h}_{ij} \in \mathbb{C}^{1 \times N_i}$ denotes the channel vector between node $i$, ($i \in \{A, B, R\}$), and node $j$, ($j \in \{B, R\}$). The channel matrix between node $i$, ($i \in \{A, B, R\}$), and the multiple colluding eavesdroppers within cluster $\mathcal{G}_j$ is denoted by $\mathbf{G}_i^j \in \mathbb{C}^{N_j^e \times N_i}$. The thermal noises at Bob, Ray and eavesdroppers are modeled as zero-mean additive white Gaussian noise (AWGN) vectors with powers $\kappa_B$, $\kappa_R$ and $\kappa_E$ Watts, respectively.

## III. SECRECY RATE OF PROPOSED SCHEME

According to the described scheme, the transmit vectors by Alice and Bob are given by

$$\mathbf{s}_A = \mathbf{a}_{\text{inf}} x + \mathbf{A}_{\text{AN}} \mathbf{z}_A \text{ and } \mathbf{s}_B = \mathbf{B}_{\text{AN}} \mathbf{z}_B, \quad (1)$$

respectively, where $\mathbf{a}_{\text{inf}} \in \mathbb{C}^{N_A \times 1}$ is the data precoding vector at Alice, $\mathbf{A}_{\text{AN}} \in \mathbb{C}^{N_A \times (N_A - 1)}$ is the precoding matrix of the AN vector $\mathbf{z}_A \in \mathbb{C}^{(N_A - 1) \times 1}$ designed to be orthogonal to the Alice-to-relay channel vector, i.e., $\mathbf{h}_{AR} \mathbf{A}_{\text{AN}} = \mathbf{0}_{1 \times (N_A - 1)}$, and $\mathbf{B}_{\text{AN}} \in \mathbb{C}^{N_B \times (N_B - 1)}$ is designed at Bob to be orthogonal to his self-interference channel vector, i.e., $\mathbf{h}_{BB} \mathbf{B}_{\text{AN}} = \mathbf{0}_{1 \times (N_B - 1)}$. Thus, the received signal at Ray is given by

$$y_R = \mathbf{h}_{AR} \mathbf{a}_{\text{inf}} x + \mathbf{h}_{BR} \mathbf{B}_{\text{AN}} \mathbf{z}_B + n_R \quad (2)$$

where $n_R \in \mathbb{C}$ is the AWGN signal at Ray's receiver. In order to maximize the SINR at Ray, the precoding vector $\mathbf{a}_{\text{inf}}$ should be equal to $\frac{\mathbf{h}_{AR}^*}{\|\mathbf{h}_{AR}\|}$. Simultaneously, Ray transmits the following signal to Bob

$$\mathbf{s}_{RB} = \mathbf{w}_R y_R + \mathbf{W}_{\text{AN}} \mathbf{z}_R \quad (3)$$

where $\mathbf{w}_R \in \mathbb{C}^{N_R \times 1}$ is the transmit precoder vector used at Ray, and $\mathbf{W}_{\text{AN}} \in \mathbb{C}^{N_R \times (N_R - 2)}$ is the precoding matrix of $\mathbf{z}_R \in \mathbb{C}^{(N_R - 2) \times 1}$ which nulls the self-interference at Ray and Bob. Thus, $\mathbf{h}_{RB} \mathbf{W}_{\text{AN}} = \mathbf{0}_{1 \times (N_R - 2)}$ and $\mathbf{h}_{RR} \mathbf{W}_{\text{AN}} = \mathbf{0}_{1 \times (N_R - 2)}$. To avoid system instability, where all the previously transmitted data symbols are amplified and forwarded over time, we have to set $\mathbf{h}_{RR} \mathbf{w}_R = 0$. Consequently, we assume zero-forcing processing for the signal transmitted by Ray at his own receiver. Under this constraint, the precoding vector $\mathbf{w}_R$ is designed to maximize the power received at Bob, while canceling out the self-interference at Ray. Following [12], $\mathbf{w}_R$ can be computed by solving the following optimization problem

$$\max_{\mathbf{w}_R} |\mathbf{h}_{RB}\mathbf{w}_R|, \text{ s.t. } |\mathbf{h}_{RR}\mathbf{w}_R| = 0, \|\mathbf{w}_R\|^2 = 1, \quad (4)$$

and is given by

$$\mathbf{w}_R = \frac{\Psi \mathbf{h}_{RB}^*}{\|\Psi \mathbf{h}_{RB}^*\|}, \quad (5)$$

where $\Psi = \mathbf{I}_{N_R} - \frac{\mathbf{h}_{RR}^* \mathbf{h}_{RR}}{\|\mathbf{h}_{RR}\|^2}$. Then, (3) can written as

$$\mathbf{s}_{RB} = \mathbf{w}_R |\mathbf{h}_{AR}|x + \mathbf{w}_R \mathbf{h}_{BR} \mathbf{B}_{AN} \mathbf{z}_B + \mathbf{w}_R n_R + \mathbf{W}_{AN} \mathbf{z}_R. \quad (6)$$

Therefore, the received signal at Bob is given by

$$y_B = \mathbf{h}_{RB}\mathbf{w}_R|\mathbf{h}_{AR}|x + \mathbf{h}_{RB}\mathbf{w}_R\mathbf{h}_{BR}\mathbf{B}_{AN}\mathbf{z}_B + \mathbf{h}_{RB}\mathbf{w}_R n_R + n_B. \quad (7)$$

Hence, the achievable rate of the legitimate system is given by

$$\mathcal{R}_{AB} = \log_2 \left( 1 + \frac{|\mathbf{h}_{RB}\mathbf{w}_R|^2 |\mathbf{h}_{AR}|^2 \theta_A P_A \theta_R P_R}{|\mathbf{h}_{RB}\mathbf{w}_R|^2 \kappa_R + \kappa_B} \right), \quad (8)$$

where $P_A$ and $P_R$ are the transmit powers at Alice and Ray, respectively. Moreover, Ray's interception rate is given by

$$\mathcal{R}_{AR} = \log_2 \det \left( 1 + \frac{|\mathbf{h}_{AR}|^2 \theta_A P_A}{|\mathbf{h}_{BR}\mathbf{B}_{AN}|^2 P_B + \kappa_R} \right). \quad (9)$$

Each alien eavesdropper receives two versions of the same symbol but shifted by one time symbol processing delay (due to delay in processing at Ray). Hence, after exchanging and combining information signals, the received signal at the eavesdroppers in cluster $\mathcal{G}_i$ is given by

$$\begin{aligned} \mathbf{y}_{\mathcal{G}_i}(t) = {} & \mathbf{G}_A^i \mathbf{a}_{\inf} x(t) + \mathbf{G}_A^i \mathbf{A}_{AN}\mathbf{z}_A(t) + \mathbf{G}_B^i \mathbf{B}_{AN}\mathbf{z}_B(t) \\ & + \mathbf{G}_R^i \mathbf{w}_R |\mathbf{h}_{AR}|x(t-1) + \mathbf{G}_R^i \mathbf{w}_R \mathbf{h}_{AR}\mathbf{A}_{AN}\mathbf{z}_A(t-1) \\ & + \mathbf{G}_R^i \mathbf{w}_R n_R(t-1) + \mathbf{G}_R^i \mathbf{W}_{AN}\mathbf{z}_R(t) \\ & + \mathbf{G}_R^i \mathbf{w}_R \mathbf{h}_{BR}\mathbf{B}_{AN}\mathbf{z}_B(t-1) + n_{E_i}(t), \end{aligned} \quad (10)$$

where $n_{E_i}$ is the AWGN signal vector at the eavesdroppers belonging to $\mathcal{G}_i$. For better data interception, each cluster of eavesdroppers needs to rearrange the received signals and decode the entire signal as a single block. Specifically, the received signal at the colluding eavesdroppers in $\mathcal{G}_i$ is

$$\begin{aligned} \tilde{\mathbf{y}}_{\mathcal{G}_i} = {} & \text{blkd}\{\mathbf{G}_A^i \mathbf{a}_{\inf}, \mathbf{G}_A^i \mathbf{a}_{\inf}, \ldots, \mathbf{G}_A^i \mathbf{a}_{\inf}\}\tilde{\mathbf{x}} \\ & + \text{blkd}\{\mathbf{G}_A^i \mathbf{A}_{AN}, \mathbf{G}_A^i \mathbf{A}_{AN}, \ldots, \mathbf{G}_A^i \mathbf{A}_{AN}\}\tilde{\mathbf{z}}_A \\ & + \text{blkd}\{\mathbf{G}_B^i \mathbf{B}_{AN}, \mathbf{G}_B^i \mathbf{B}_{AN}, \ldots, \mathbf{G}_B^i \mathbf{B}_{AN}\}\tilde{\mathbf{z}}_B \\ & + \text{blkd}\{\mathbf{G}_R^i \mathbf{w}_R |\mathbf{h}_{AR}|, \ldots, \mathbf{G}_R^i \mathbf{w}_R |\mathbf{h}_{AR}|\}\mathbf{S}\tilde{\mathbf{x}} \\ & + \text{blkd}\{\mathbf{G}_R^i \mathbf{w}_R \mathbf{h}_{AR}\mathbf{A}_{AN}, \ldots, \mathbf{G}_R^i \mathbf{w}_R \mathbf{h}_{AR}\mathbf{A}_{AN}\}\mathbf{S}\tilde{\mathbf{z}}_A \\ & + \text{blkd}\{\mathbf{G}_R^i \mathbf{w}_R, \mathbf{G}_R^i \mathbf{w}_R, \ldots, \mathbf{G}_R^i \mathbf{w}_R\}\tilde{\mathbf{n}}_R \\ & + \text{blkd}\{\mathbf{G}_R^i \mathbf{W}_{AN}, \ldots, \mathbf{G}_R^i \mathbf{W}_{AN}\}\tilde{\mathbf{z}}_R \\ & + \text{blkd}\{\mathbf{G}_R^i \mathbf{w}_R \mathbf{h}_{BR}\mathbf{B}_{AN}, \ldots, \mathbf{G}_R^i \mathbf{w}_R \mathbf{h}_{BR}\mathbf{B}_{AN}\}\mathbf{S}\tilde{\mathbf{z}}_B + \tilde{\mathbf{n}}_{E_i}, \end{aligned} \quad (11)$$

where $\mathbf{S} = [\mathbf{0}_{1\times\mathcal{B}}; [\mathbf{I}_{(\mathcal{B}-1)}]_{1:\mathcal{B}}]$ is a matrix with an all-zero vector as its first row and the rows of the identity matrix on the other rows, $\mathcal{B} = \lfloor \mathcal{W}T \rfloor$ is the block size of the codeword, $\tilde{\mathbf{x}} = [x(0), x(1), \ldots, x(\mathcal{B})]$, $\tilde{\mathbf{z}}_A = [\mathbf{z}_A(0), \ldots, \mathbf{z}_A(\mathcal{B})]$ with $x(0) = 0$, $\tilde{\mathbf{n}}_R = [n_R(0), n_R(1), \ldots, n_R(\mathcal{B})]$, $\tilde{\mathbf{z}}_R = [\mathbf{z}_R(0), \ldots, \mathbf{z}_R(\mathcal{B})]$ and $\tilde{\mathbf{z}}_B = [\mathbf{z}_B(0), \ldots, \mathbf{z}_B(\mathcal{B})]$. The signal vector in (11) can be rewritten as

$$\begin{aligned} \tilde{\mathbf{y}}_{\mathcal{G}_i} = {} & \mathbf{D}\tilde{\mathbf{x}} + \mathbf{C}_A\tilde{\mathbf{z}}_A + \mathbf{C}_R\tilde{\mathbf{z}}_R + \mathbf{C}_B\tilde{\mathbf{z}}_B \\ & + \text{blkd}\{\mathbf{G}_R^i\mathbf{w}_R, \ldots, \mathbf{G}_R^i\mathbf{w}_R\}\tilde{\mathbf{n}}_R + \tilde{\mathbf{n}}_{E_i}, \end{aligned} \quad (12)$$

where $\mathbf{D}$, $\mathbf{C}_A$, $\mathbf{C}_R$ and $\mathbf{C}_B$ are given in (13) at the top of the next page.

Based on (8), (9) and (14), the achievable secrecy rate for the proposed scheme is given by

$$\mathcal{R}_{\sec} = \left[ \mathcal{R}_{AB} - \max\{\mathcal{R}_{AR}, \max_{1 \leq i \leq M} \mathcal{R}_{\mathcal{G}_i}\} \right]^+. \quad (15)$$

We assume that Ray and the different clusters of eavesdroppers process their received signals independently and only eavesdroppers in the same cluster collude (i.e., jointly process their received signals).

## IV. SIMULATION RESULTS

In this section, we provide simulation results to evaluate the achieved secrecy rate of our proposed scheme in a Rayleigh fading environment where the channel coefficients are circularly-symmetric Gaussian random variables with zero mean and unit variance. We set the transmit powers of Alice, Bob and Ray to $P_A = P_B = P_R = 10$ dBm. The AWGN power at any receiving node is normalized to 0 dBm. We assume that each transmission block is associated with a normalized time slot of duration $T = 10$ milliseconds and bandwidth $W = 1$ MHz. The number of transmit antennas at Alice, Bob and Ray are $N_A = N_B = N_R = 4$. We assume that multiple single-antenna eavesdroppers are grouped into one cluster. Then, the total number of antennas, denoted as $N_E$, in the cluster is equal to the number of nodes it contains.

Figs. 2 and 3 show the achievable secrecy rate of our proposed scheme versus the data power allocation factors, $\theta_A$ and $\theta_R$, at Alice and Ray, respectively, for different total number of receive antennas, $N_E$, of the eavesdropping cluster. We assumed: $N_E = 4$ and $N_E = 2$. In both scenarios, for fixed $\theta_A$ and $\theta_R$, as $N_E$ increases, the average secrecy rate decreases. Obviously, the eavesdropping cluster's ability to decode the information increases with the total number of its antennas.

It is also clear that a dynamic power allocation increases the secrecy rate of our proposed scheme. As shown in Fig. 3, the optimal values of $\theta_A$ and $\theta_R$ are close to 1. The reason is that when $N_E$ is small, the eavesdropping nodes, even with collusion, do not have enough ability to decode the received signals that are combined with Bob's jamming signal. In this case, the untrusted relay with a single receive antenna will be more powerful to eavesdrop the data signal. Hence, in such cases, Bob's jamming signal is enough to achieve secure transmission and Alice and Ray should allocate more power to transmit data signals to increase the achievable rate at Bob.

However in Fig. 2, when $N_E = 4$, the eavesdropping cluster is more powerful than the untrusted relay and becomes a threat to secrecy transmission. Thus, the eavesdropping SINR dominates in the secrecy rate expression. Therefore, Alice and Ray should carefully use some of their available transmit powers to send jamming signals. As shown in Fig. 2, the optimal values of $\theta_A$ and $\theta_R$ are between 40% and 75% and between 80% and 100%, respectively. Furthermore, we can

$$\begin{aligned}
\mathbf{D} &= \text{blkd}\{\mathbf{G}_{\text{A}}^{\text{i}}\mathbf{a}_{\text{inf}},\ldots,\mathbf{G}_{\text{A}}^{\text{i}}\mathbf{a}_{\text{inf}}\} + \text{blkd}\{\mathbf{G}_{\text{R}}^{\text{i}}\mathbf{w}_{\text{R}}|\mathbf{h}_{\text{AR}}|,\ldots,\mathbf{G}_{\text{R}}^{\text{i}}\mathbf{w}_{\text{R}}|\mathbf{h}_{\text{AR}}|\}\mathbf{S},\\
\mathbf{C}_A &= \text{blkd}\{\mathbf{G}_{\text{A}}^{\text{i}}\mathbf{A}_{\text{AN}},\ldots,\mathbf{G}_{\text{A}}^{\text{i}}\mathbf{A}_{\text{AN}}\} + \text{blkd}\{\mathbf{G}_{\text{R}}^{\text{i}}\mathbf{w}_{\text{R}}\mathbf{h}_{\text{AR}}\mathbf{A}_{\text{AN}},\ldots,\mathbf{G}_{\text{R}}^{\text{i}}\mathbf{w}_{\text{R}}\mathbf{h}_{\text{AR}}\mathbf{A}_{\text{AN}}\}\mathbf{S},\\
\mathbf{C}_R &= \text{blkd}\{\mathbf{G}_{\text{R}}^{\text{i}}\mathbf{W}_{\text{AN}},\ldots,\mathbf{G}_{\text{R}}^{\text{i}}\mathbf{W}_{\text{AN}}\},\\
\mathbf{C}_B &= \text{blkd}\{\mathbf{G}_{\text{B}}^{\text{i}}\mathbf{B}_{\text{AN}},\ldots,\mathbf{G}_{\text{B}}^{\text{i}}\mathbf{B}_{\text{AN}}\} + \text{blkd}\{\mathbf{G}_{\text{R}}^{\text{i}}\mathbf{w}_{\text{R}}\mathbf{h}_{\text{BR}}\mathbf{B}_{\text{AN}},\ldots,\mathbf{G}_{\text{R}}^{\text{i}}\mathbf{w}_{\text{R}}\mathbf{h}_{\text{BR}}\mathbf{B}_{\text{AN}}\}.
\end{aligned} \quad (13)$$

$$\mathcal{R}_{\mathcal{G}_i} = \frac{1}{\mathcal{B}}\log_2 \det\left(\mathbf{I}_\mathcal{B} + \theta_R P_A \mathbf{D}\mathbf{D}^*\left(\overline{\theta_A}P_A\mathbf{C}_A\mathbf{C}_A^* + \overline{\theta_R}P_R\mathbf{C}_R\mathbf{C}_R^* + P_B\mathbf{C}_B\mathbf{C}_B^* + \kappa_E\left(\mathbf{I}_\mathcal{B} + \text{blkd}\{\mathbf{G}_{\text{R}}^{\text{i}}\mathbf{w}_{\text{R}},\ldots,\mathbf{G}_{\text{R}}^{\text{i}}\mathbf{w}_{\text{R}}\}\text{blkd}\{\mathbf{G}_{\text{R}}^{\text{i}}\mathbf{w}_{\text{R}},\ldots,\mathbf{G}_{\text{R}}^{\text{i}}\mathbf{w}_{\text{R}}\}^*\right)\right)^{-1}\right). \quad (14)$$

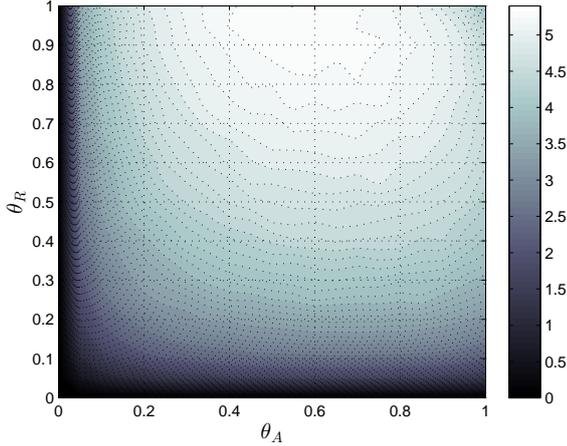

Fig. 2. The achievable secrecy rate versus $\theta_A$ and $\theta_R$ when $N_E = 4$.

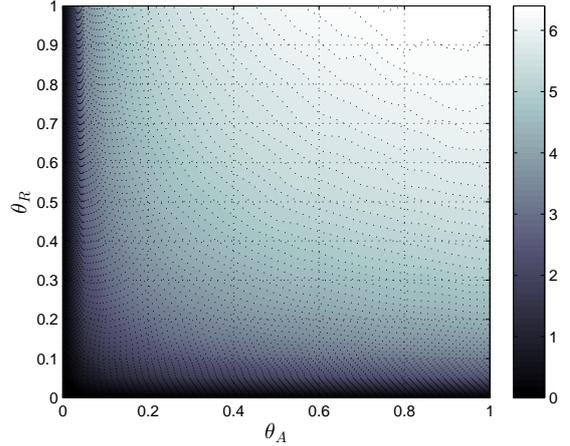

Fig. 3. The achievable secrecy rate versus $\theta_A$ and $\theta_R$ when $N_E = 2$.

see that Alice should expend more jamming power than the untrusted relay. In fact, the noise signal and Bob's jamming signal are amplified at Ray. Hence, when $\theta_R$ increases, the jamming noise at the eavesdropping cluster increases while Bob's rate increases as well. On the other hand, along with DBJ, transmitting AN from Alice, i.e., $\theta_A \neq 1$, is needed to protect her own transmissions.

## V. CONCLUSION

We analyzed the PLS of FD untrusted relay networks in the presence of multiple external clusters of colluding eavesdroppers. We proposed an AN-aided secure transmission scheme in which the untrusted relay and the destination cooperate with the source to confound the eavesdroppers and the relay, respectively, by generating AN signals. Our numerical results quantified the secrecy rate gains of the AN and the negative impact of increasing the total number of antennas of the eavesdroppers clusters. We showed that the optimal values of the power factor at Alice, $\theta_A$, and the power factor at the relay, $\theta_R$, are between 40% and 75% and between 80% and 100%, respectively. When the total number of receive antennas in the eavesdropping clusters is small, almost all the power should be allocated to the information signal at both the source and the relay. However, more jamming power is required as the number of the eavesdropping cluster's antennas increases.